# The Nonlinear Asymptotic Stage of the Rayleigh-Taylor Instability with Wide Bubbles and Narrowing Spikes

V. M. CHERNIAVSKI AND YU. M. SHTEMLER

*Institute of New Technologies, Moscow, Russia*

The potential flow of an incompressible inviscid heavy fluid over a light one is considered. The integral version of the method of matched asymptotic expansion is applied to the construction of the solution over long intervals of time. The asymptotic solution describes the flow in which a bubble rises with constant speed and the "tongue" is in free fall. The outer expansion is stationary, but the inner one depends on time.

It is shown that the solution exists within the same range of Froude number obtained previously by Vanden-Broeck (1984a,b). The Froude number and the solution depend on the initial energy of the disturbance. At the top of the bubble, the derivative of the free-surface curvature has a discontinuity when the Froude number is not equal to 0.23. This makes it possible to identify the choice of the solution obtained in a number of studies with the presence of an artificial numerical surface tension. The first correction term in the neighborhood of the tongue is obtained when large surface tension is included.

## INTRODUCTION

For two-layered fluid media over long intervals of time, rising bubbles and falling tongues have been observed in a number of laboratory experiments [Emmons, 1960]. It is characterized by a constant dimensionless speed of a wide bubble (Froude number $Fr = 0.2 - 0.3$) and a constant acceleration of a narrowing tongue.

As usual, the stationary solution is represented as a sum of a singular and a regular series [Birkhoff and Carter, 1957]. Birkhoff and Carter (1957) and Vanden-Broeck (1984a,b) have lost the second term in the singular series. To obtain the correct (nonoscillatory) solution and to preserve the energy conservation law, it is necessary to keep this second term.

Instead of the inner expansion, we use the conditions at the jump between the branches of the steady solution. These conditions are obtained using the conservation laws, which give the correct asymptotic formulas for a steady solution. In addition, the outcome of these are employed to estimate numerical errors. We show that the solution exists within the same range of Froude number ($Fr$) as obtained previously [Vanden-Broeck, 1984a]. We find that $Fr$ depends on the initial energy of the disturbance. At the top of the bubble, the derivative of the free-surface curvature has a discontinuity when $Fr \neq 0.23$. This makes it possible to identify the choice of solution obtained in a number of studies with the presence of an artificial numerical surface tension. We obtain the first correction term in the neighborhood of the tongue when large surface tension is included.

## FORMULATION OF THE PROBLEM

The potential flow of an inviscid incompressible heavy fluid lying above a light one is investigated. In a Cartesian coordinate system $XY$, let the plane $Y = 0$ be the unperturbed free surface of a heavy liquid ($Y \geq 0$). Let us consider a two-dimensional potential flow of the fluid with unit density. We assume at long times $T$ that the flow is spatially periodic in $X$ and symmetric with respect to the $Y$ axis

$$\Delta \Phi = 0,$$
$$\Phi(X + \lambda, Y, T) = \Phi(X, Y, T),$$
$$\Phi(-X, Y, T) = \Phi(X, Y, T),$$
$$\frac{d\Gamma}{dT} = 0 \quad (\Gamma = 0),$$
$$\Gamma(X + \lambda, Y, T) = \Gamma(X, Y, T),$$
$$\Gamma(-X, Y, T) = \Gamma(X, Y, T),$$
$$P = 0 \quad (\Gamma = 0), \tag{1}$$

$$P = -\left(Y + \frac{1}{2}\left|\frac{dF}{dZ}\right|^2 + \frac{\partial \Phi}{\partial T}\right),$$

$$Z = X + iY, \quad F = \Phi + i\Psi,$$

$$\left|\frac{dF}{dZ}\right| \to 0, \quad (Y \to \infty) \ .$$

Here $P$ is the pressure, $F$ is the complex potential, $\Delta$ is the Laplacian operator, and $\Gamma(X, Y, T) = 0$ is the equation of the free surface. The pressure outside the liquid is assumed equal to zero. To transform to dimensionless variables, we shall take the wavelength $\lambda$ and the free fall acceleration $g$ as characteristic scales. The flow is shown schematically in Figure 1. (The dashed curves show the time evolution of the free surface.) We assume that the bubbles rise when $X = \pm 1/2$, while a tongue of heavy liquid collapses when $X = 0$. Assuming that the coordinate system moves with the velocity of the bubble top, $w = Fr$, we have

$$T = t - t_0, \quad Z = iwt + z + iy_0,$$
$$F(Z, T) = -iwz + f(z, t) + \Phi_0(t),$$
$$P(X, Y, T) = p(x, y, t), \tag{2}$$
$$\Gamma(X, Y, T) = \gamma(x, y, t),$$
$$z = x + iy, \quad f = \phi + i\psi \ .$$

Here, $\Phi_0(t) = -w(t^2 - wt)/2 - y_0 t + \phi_0$. We assume that the initially unknown constant $w_0$ is the velocity of the bubble top, when $t$ tends to infinity. We define the dimensionless value $w$ in terms of the dimensional velocity $w_0$ as

$$w = \frac{w_0}{\sqrt{\lambda g}} \ .$$

We now transform one period of the flow on the $z$-plane onto a halfstrip on the $q$-plane

$$q = s + ir, \quad r > 0, \quad |s| \le \pi,$$
$$\frac{\partial x}{\partial s} = \frac{\partial y}{\partial r}, \quad \frac{\partial x}{\partial r} = -\frac{\partial y}{\partial s},$$
$$x(-s, r, t) = -x(s, r, t), \quad x(\pi, r, t) = \frac{1}{2}, \tag{3}$$
$$\gamma(x(s, 0, t), y(s, 0, y), t) = 0 \quad |s| \le \pi,$$
$$\frac{\partial y}{\partial r} \to \frac{\pi}{2}, \quad r \to \infty \ .$$

### THE FORM OF THE STEADY-STATE SOLUTION IN THE ABSENCE OF SURFACE TENSION

Let us consider the steady-state solution $f$, $z$ of the system of equations (1)–(3), which is an outer solution in the asymptotic expansion in $1/t$ which is valid outside the region of the narrow tongue. According to Birkhoff and Carter [1957] we have

$$f = \frac{w}{2\pi} \ln \frac{\theta}{(1-\theta)^2} - wy_\infty \tag{4}$$

$$\theta = \exp(iq), \quad y_\infty = \text{const}.$$

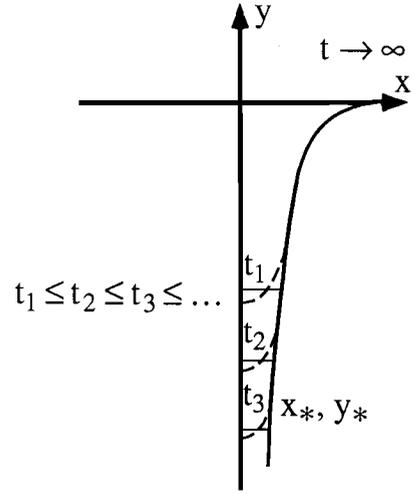

Figure 1. The flow is shown schematically. The dashed curves show the time evolution of the free surface. The bubbles rise at $X = \pm 0.5$, the tongue of heavy liquid collapses at $X = 0$.

The free surface is described by a curve which everywhere is analytic, except at points where the velocity is infinite ($\theta = 1$) and, possibly, at points with zero velocity ($\theta = -1$) [Birkhoff, 1957].

The unknown $z$ is sought in the form

$$iz = L^{2/3} \sum_{k=0}^{n} D_k L^{-k} + \sum_{k=0}^{N} iz_k \theta_k + \frac{i}{2\pi} q$$
$$\left(L = \ln\left(\frac{1-\theta}{2}\right)^2 - 1\right) \ . \tag{5}$$

This solution includes $D_1$ and differs from Birkhoff (1957) and Vanden-Broeck (1984a). The coefficients $D_k$ for $k = 1, 2, \cdots$ are determined by using an asymptotic expansion procedure in powers of $L$ as $\theta \to 1$, and $D_0 = (9w^2/32\pi^2)^{1/3}$, while $z_k$ is found from the dynamic condition on the free surface with the aid of the Fast Fourier Transform and Newton's method.

### CONSERVATION LAWS IN THE ABSENCE OF FRICTION

Due to the symmetry of the problem and the conditions at infinity, the conservation laws for the system of equations (1) and (2) can be written in the form

$$\frac{di_1}{dt} = -\frac{w}{2}\phi_\infty, \quad i_1 = \frac{1}{2}\int_S \left(\phi d\psi - y^2 dx\right)$$
$$\frac{di_2}{dt} = w, \quad i_2 = -\int_S y dx,$$
$$\frac{di_3}{dt} = -i_2 - \frac{w^2}{2}, \quad i_3 = -\int_S \phi dx,$$
$$\frac{di_4}{dt} = i_3 + \phi_\infty, \quad i_4 = -\frac{1}{2}\int_S y^2 dx, \tag{6}$$

$$\frac{di_5}{dt} = 4i_1 - 7i_4 + b(t), \quad i_5 = \int_S \phi(ydx - xdy),$$

$$b = \frac{1}{2} \int_h^\infty (\phi_y^2 - w^2)\big|_{s=\pi} dy + \phi_\pi h_t - \frac{1}{2}h^2 - \frac{1}{2}w^2 h .$$

Here, $h(t)$ is the ordinate of the bubble top with respect to the moving coordinate system, $\phi_\infty(t)$ is an arbitrary function depending on time. The integrals $i_1, i_2, \cdots i_5$ represent the deviations of the total energy, mass, momentum, potential energy, and "virial" from the corresponding values in the unperturbed state $\phi_\pi = \phi(\pi, 0, t)$.

Let us examine the behavior of the nonstationary solution as it approaches the stationary solution, equations (4) and (5), outside the narrow tongue region. It can be shown, up to exponentially small terms and parameters $y_0$ and $\phi_0$, that

$$\phi_\infty(t) \to 0, \quad h(t) \to 0, \quad dh/dt \to 0,$$
$$b(t) \to \int_0^\infty (\phi_y^2 - w^2)\big|_{s=\pi} dy \quad (t \to \infty) . \quad (7)$$

The order of the nonstationary solution in the neighborhood of the tongue is obtained from Eq. (6) using Eq. (7), namely

$$x \sim t^{-1}, \quad y \sim t^2, \quad \phi \sim t^3, \quad \psi \sim 1 \quad (t \to \infty) . \quad (8)$$

## NONSTATIONARY JUMP CONDITIONS IN THE NEIGHBORHOOD OF THE TONGUE

The solution in the neighborhood of the tongue can be described by an integral along a contour which approximates the free surface. (In Figure 1, the smooth curve corresponds to the stationary solution, while the dot-dashed curves represent the nonstationary jump; $x_*(t)$ and $y_*(t)$ are the coordinates of the right side of the jump at the time $t$). The even functions are discontinuous and the odd ones are continuous:

$$y(s, 0) = y_*(x_*), \quad x(s, 0) = \pm x_*,$$
$$p(s, 0) = p_*(x_*), \quad \phi(s, 0) = \phi_*(x_*), \quad (9)$$
$$\psi(s, 0) = \pm \psi_*, \quad s = \pm s_* .$$

For $r = 0$ and $s \to 0$ $(t \to \infty)$ from Eqs. (4), (5) and (8) we obtain

$$y_*(x_*) = -\frac{w^2}{8x_*^2} + O(x_*^4),$$

$$p_*(x_*) = O(x_*^4),$$

$$\phi_*(x_*) = \left(\frac{w}{x_*}\right)^3 \frac{1}{24} + 2\frac{k}{w} + O(x_*^3), \quad (10)$$

$$\psi_*(x_*) = \frac{w}{2} + O(x_*^6),$$

$$x_*(t) = \frac{\mu}{t} .$$

## CONSEQUENCES OF THE CONSERVATION LAWS FOR THE STATIONARY CASE

The proportionality constant $\mu$ is determined by the conservation laws. Using Eqs. (9) and (10), we evaluate the integrals $i_k$ in Eq. (6) for the limiting solution of the nonstationary problem when $t \to \infty$. The solution is determined by solving Eq. (5) for the stationary problem and Eqs. (9) and (10), namely

$$i_1 = v + k, \quad i_2 = m + \frac{w^2}{2x_*},$$
$$i_3 = j - \frac{(w/2)^3}{x_*^2}, \quad i_4 = v - \frac{(w/2)^4}{3x_*^3}, \quad (11)$$
$$i_5 = d + \frac{7(w/2)^5}{12x_*^4} + \frac{2kw}{x_*} .$$

Here, $v + k$, $m$, $j$, $v$ and $d$ are the regular parts of the integrals $i_k$ with $k = 1, \cdots, 5$, and

$$m = -2 \int_0^{1/2} [y - y_*(x)] dx - \frac{w^2}{2} + O(x_*^5),$$

$$v = -\int_0^{1/2} [y^2 - y_*^2(x)] dx - \frac{w^4}{24} + O(x_*^3),$$

$$j = -2 \int_0^{1/2} [\phi - \phi_*(x)] dx + \frac{w^3}{6} - 2\frac{k}{w} + O(x_*^4),$$

$$k = -\frac{w^2}{2}\left(y_\infty + \frac{L_\infty - 3D_1/2}{2\pi}\right),$$

$$y_\infty = z_0 + \sum_{l=0} D_l L_\infty^{2/3-l},$$

$$L_\infty = 1 + \ln 4.$$

Integrating Eq. (6) with respect to time and taking into consideration Eq. (11), we obtain:

$$\mu = \frac{w}{2} \quad (12)$$

$$m = -\frac{w^2}{2}, \quad j = 0, \quad v = \frac{b_c}{3} \quad (13)$$

$$c_1 = v + k, \quad c_2 = -\frac{w^2}{2}, \quad c_3 = 0 . \quad (14)$$

Here the $c_k$, with $k = 1, 2, 3$, are the initial values of the $i_k$ at $t = 0$. According to Eqs. (6) and (7), the integrals $i_4$ and $i_5$ depend on the transition process in the first order approximation, and the equations for them have not been expressed in this approximation. According to Eq. (10), the value of $\mu$ in Eq. (12) determines the average acceleration of the tongue, which equals the acceleration of free fall. The integral identities (13) are consequences of the conservation laws for the stationary case. Equations (14) determine the particular initial data corresponding to the limiting solution.

## FROUDE NUMBER DEPENDENCE ON THE INITIAL TOTAL ENERGY

From Eqs. (6) and (14), we obtain the dependence of the initial data $c_k^0 = i_k(t_0, c_n)$ on the parameter $t_0$:

$$c_1^0 + \frac{w}{2}\phi_\infty(t_0) = v + k,$$
$$c_2^0 = wt_0 - \frac{w^2}{2}, \quad c_3^0 = -\frac{wt_0^2}{2} \quad (15)$$

For comparison with physical experiments and numerical simulations, we can use Eq. (2) to obtain the relationship between the initial data with respect to the moving and fixed coordinate system ($t = t_0$):

$$E = c_1^0 - w\phi_\infty/2 + c_2^0(Y_0 + w^2/2) + wc_3^0 - Y_0^2/2,$$
$$M = c_2^0 - Y_0, \quad J = c_3^0 + wc_2^0 - \Phi_0. \quad (16)$$

Here $Y_0 = y_0 + wt$; $E$, $M$ and $J$ are the initial deviations in the total energy, mass and momentum with respect to the fixed coordinate system. These deviations are obtained by formal replacement of $x$, $y$, $\phi$, and $\psi$ by $X$, $Y$, $\Phi$ and $\Psi$ for the $i_k$ defined in Eq. (6). Without loss of generality we set $M = 0$ and $J = 0$ and, thereby choose a system of coordinates and the constant for the potential. On eliminating $c_k^0$ from Eqs. (15) and (16), we obtain the dependence of the parameters $w$, $y_0$ and $\phi_0$ on the initial data, namely

$$E = v(w) + k(w) - \frac{w^3}{8},$$
$$y_0 = -\frac{w^2}{2}, \quad \phi_0 = -\frac{w^3}{2}. \quad (17)$$

## CALCULATION RESULTS

As a result of the calculations, it was found that a surface with a smooth curvature exists only for a certain value of $D_1 = D_*$. When $D_1 \neq D_*$, the solution oscillates and the identities (13) are no longer satisfied. A stationary solution corresponding to $D_*$ has been found for Froude numbers in the range $0 < w < 0.37$. This interval is approximately the same as that has been found previously [Vanden-Broeck, 1984a]. Figure 2 shows the shape of the free surface $y = y(x)$, for three values of $w$. The derivative of the curvature of the interface surface at the top of a bubble has a discontinuity over the entire range of Froude numbers, except at one value ($w = 0.23$). Figure 3 shows the curvature $K = K(x)$ for three values of $w$.

We find that two possible values for the velocity of a rising bubble can exist for a single initial energy. Figure 4 shows the function $w(E)$ given by Eq. (17).

## SMALL AMOUNT OF SURFACE TENSION AND DISCONTINUITY OF THE CURVATURE DERIVATIVE

When the surface tension with coefficient $\sigma$ is taken into account, the dynamic condition at the free surface $y = y(x)$ has the form

$$y(x) + \frac{1}{2}\left|\frac{f}{z}\right|^2 + \sigma K = \sigma K_0,$$

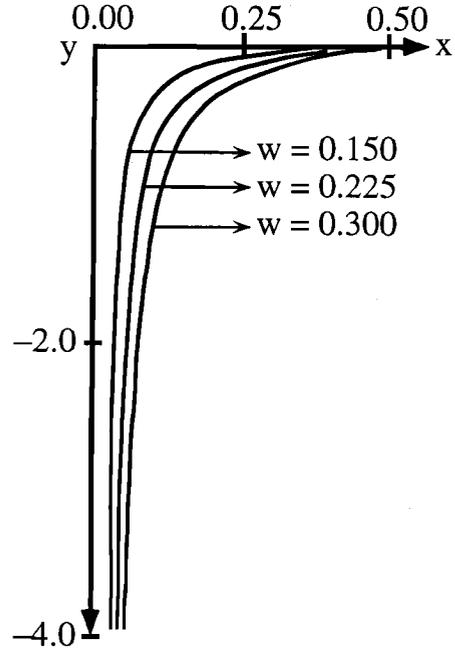

Figure 2. The shape of the free surface $Y = y(x)$ for three values of the Froude number, namely $w = 0.150, 0.225$, and $0.300$.

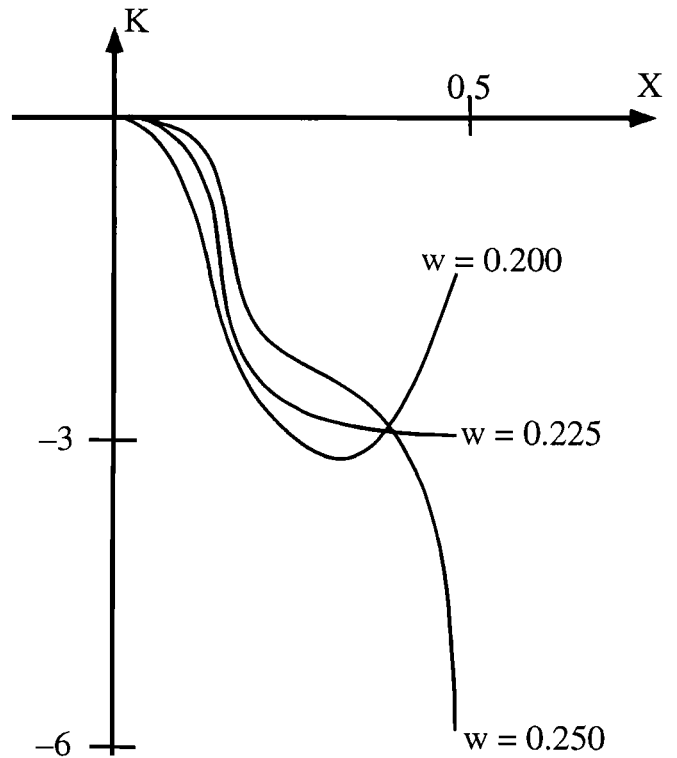

Figure 3. The curvature $K = K(x)$ for three values of the Froude number, namely $w = 0.200, 0.225, 0.250$.

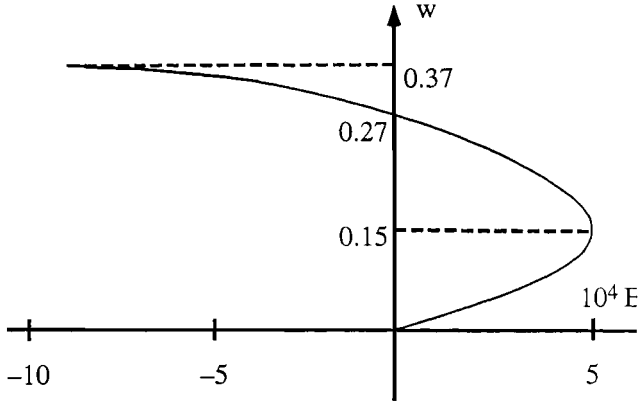

Figure 4. The dependence of Froude number upon the initial energy of the disturbances $w = w(E)$ given by Eq. (17).

$$f_z = \frac{df}{dz}, \quad K(x) = \frac{d}{dx}\left(\frac{y_x}{\sqrt{1+y_x^2}}\right), \qquad (18)$$

$$K_0 = K\left(\frac{1}{2}\right), \quad y_x = \frac{dy}{dx}.$$

If $\sigma \neq 0$ the contact angle at the top of a bubble at the wall must be specified. Let

$$\frac{dy(x)}{dx} = 0 \quad \left(x = \pm\frac{1}{2}\right). \qquad (19)$$

We seek a solution of Eq. (18) in the form

$$y(x) = y^0(x) + \sigma y^1(x) + \cdots,$$
$$f_z = f_z^0 + \sigma f_z^1 + \cdots,$$
$$K = K^0 + \cdots \quad . \qquad (20)$$

It can be shown that the boundary condition (19) is satisfied in the zeroth order of $\sigma$ and is violated in the next order, since $y_x^1 = K_x^0$ has a discontinuity at the top of the bubble for all $w$ except $w = 0.23$.

A small amount of surface tension may, therefore, isolate a fully defined limiting velocity at which the bubble will rise and, thus, explain the results obtained by Baker et al. (1980).

### THE FIRST CORRECTION TERM DUE TO A LARGE SURFACE TENSION COEFFICIENT

When the surface tension coefficient $\sigma$ is not small, the first correction term in the steady-state solution of Eq. (18) is also proportional to $\sigma$. It can be seen from the conservation laws that if the surface tension is taken into account it is necessary to add the term $-\sigma \int_S \left(\sqrt{1+y_x^2} - 1\right) dx$ to the total energy $i_1$ and to exclude the law for the "virial"—in Eq. (6) [Benjamin, 1982].

Let us take only the first term in Eq. (5), namely

$$iz^0 = D_0 L^{2/3} \quad .$$

It is easy to see that kinetic and potential energies in $i_1$ are of order $t^3$ and that the newly added term is of order $t^2$. Thus the term $iz^0$ is the main term in asymptotic expansion in the neighborhood of the tongue and the expansion (2) is also correct when $\sigma$ is not small.

It can be shown that the linearized dynamic condition (18) has the form

$$\Re\left\{iz^1 + 2\frac{d(iz^1)}{d\ln(iz^0)}\right\} = \sigma K^0 - \sigma K_0^0 \quad . \qquad (21)$$

This means that the real part of a harmonic function, say $f(q)$, is equal on the surface to $\sigma K^0 - \sigma K_0^0$. Here $K^0$ is the curvature of the zeroth order surface. Let us map the half plane $q = s + ir, r \geq 0$ onto the circle $\omega = \rho \exp(i\alpha), \rho \leq 1, q = i\frac{1-\omega}{1+\omega}$ and include the function $L_0(\omega) = \ln\frac{1+\omega}{1-\omega}$ taking the branch

$$L_0\left(e^{i\alpha}\right) = \ln\left|\cot\frac{\alpha}{2}\right| + i\frac{\pi}{2}\text{sign}(\alpha) \quad .$$

It is evident that the function $L(q)$ in Eq. (5) is equal to $-2L_0(\omega)$ with asymptotic accuracy.

We construct a Schwarz function $f(q) = F(\omega)$ using the next theorem [Pyckteev, 1982]. If the density $f(s) = F(e^{i\alpha})$ is an even function of $\alpha$ and $|\alpha| \leq \pi$, then let us select the function which is equal to density for $0 < \alpha \leq \pi$, namely

$$\phi(\zeta) = \phi_1(\zeta) + \phi_2(\zeta),$$
$$\phi(\zeta) = F(e^{i\alpha}), \quad \zeta = e^{i\alpha},$$

where $\phi_1$ is an even function, and $\phi_2$ is an odd one, for $|\alpha| \leq \pi$. Then, the Schwarz function is

$$F(\omega) = \phi_1(\omega) - i\frac{2}{\pi}\phi_2(\omega)L_0(\omega) \quad .$$

It is observed from Eq. (21) that

$$\phi_1(\zeta) = -\sigma K_0^0,$$

$$\phi_2(\zeta) = \frac{\sigma}{iz_q^0}\frac{1}{2i}\left\{\frac{d}{dq}\ln(iz_q^0) - \left[\frac{d}{dq}\ln(iz_q^0)\right]^*\right\}$$

and it is easy to show that, with asymptotic accuracy,

$$\phi_2(\omega) = \frac{\sigma}{4iD_0}2^{-2/3}L_0^{-5/3}(\omega)\left[L_0(\omega) - L_0\left(\frac{1}{w}\right)\right],$$

$$F(\omega) = -\frac{\sigma}{4\pi D_0}[2L_0(\omega)]^{-2/3}\left[L_0(\omega) - \bar{L}_0(\omega)\right] - \sigma K_0^0,$$

$$\bar{L}_0(\omega) = L_0\left(\frac{1}{\omega}\right),$$

$$f(q) = \frac{\sigma}{4\pi D_0}[2L]^{-2/3}(q)[L(q) - \bar{L}(q)] - \sigma K_0^0,$$

$$\bar{L}(q) = -2\bar{L}_0(\omega),$$

here the symbol * defines complex conjugation. Returning to Eq. (21), we have

$$\frac{1}{2}f(q) = \frac{d(iz^1)}{d\ln(iz^0)} + \frac{1}{2}iz^1$$

$$= \frac{\sigma}{2}\left[\frac{1}{4\pi D_0}L^{-2/3}(L - \tilde{L}) - K_0^0\right] \quad . \tag{22}$$

From the ordinary differential equation (22) we obtain

$$iz^1(q) = \frac{1}{\sqrt{iz^0}}\frac{1}{2}\int \frac{iz_q^0}{\sqrt{iz^0}}f(q)dq$$

and on the free surface we have

$$iz^1(s) = -i\frac{\sigma}{2D_0}L^{-2/3}(s)\,\text{sign}\,(s) - \sigma K_0^0 \quad .$$

So, the first correction term is obtained for the solution of the problem which accounts the surface tension.

### DISCUSSION

An important issue is the suitability of this two-dimensional model of flow to the real three-dimensional one. The nonlinear problem requires special treatment for three-dimensional flow and does not reduce to the two-dimensional one, as in the linear case. We have investigated this in previous work [Gertsenshtein and Cherniavskii, 1985] and our results have shown that the solution does depend on the initial values and, in particular, on some aspects of three-dimensionality. But the two-dimensional case is justified because it survives under some conditions within the framework of the three-dimensional problem. On the other hand the results of the present work demonstrate that the two-dimensional solution discussed here can exist only if the initial values have the energy within some range. So, we have considered the solution to the three-dimensional problem which satisfies a special class of initial conditions and survives under some requirements.

A second issue which emerges is the relevance of these simulations to natural phenomena. Flows which are similar to our solutions have been observed in a set of laboratory experiments— see Emmons (1960), for example. Establishing the kind of initial conditions that can occur in nature is beyond our present considerations, but should be explored in the future. However, we believe that the determination of the precise solution to this nonlinear problem is a step toward understanding such natural phenomena as mantle density differentiation governed by gravity forces. This model could be regarded as a possible process in the mantle.

### CONCLUSIONS

Asymptotic solutions which describe the behavior of a rising bubble of light fluid and of a narrow, falling tongue of heavy fluid over long times for development of the Rayleigh-Taylor instability have been obtained. It has been shown that the limiting upward velocity of a bubble (the Froude number) depends on the initial energy of the perturbations. The derivative of the curvature of the interface at the top of the bubble is discontinuous for all values except one, corresponding to a special value of the Froude number.

The first correction term in the neighborhood of the tongue has been obtained when large surface tension was taken into account. The regular part of stationary solution was obtained by numerical procedure using the Fast Fourier Transform and Newton's method.

V. M. Cherniavski and Yu. M. Shtemler, Institute of New Technologies, Kirovogradskaya 11, Moscow, 113587, Russia